\documentclass[twocolumn,dvipsnames,prx,nofootinbib,floatfix,superscriptaddress]{revtex4-2}

\usepackage{xcolor}
\usepackage{amsmath,amssymb}
\usepackage{appendix}
\usepackage{makecell}%To keep spacing of text in tables
\usepackage{relsize}
\usepackage{graphicx}
\usepackage{stmaryrd}
\usepackage{bm}
\usepackage{lipsum}
\usepackage{multirow}
\usepackage{physics}
\usepackage{float}

\usepackage{lineno}

\setcellgapes{4pt}%parameter for the spacing
\usepackage[caption=false, position=top]{subfig}

\PassOptionsToPackage{colorlinks,plainpages}{hyperref}
\usepackage[colorlinks,plainpages]{hyperref}
\hypersetup{urlcolor=blue,citecolor=blue, linkcolor=blue,hypertexnames=false}

\usepackage{xspace}
\usepackage{setspace}
\usepackage{bm}

\usepackage{orcidlink}

\newcommand{\ULNonGRTVSTOld}{\ensuremath{3.9\times 10^{-9}}}
\newcommand{\ULNonGRTVSVOld}{\ensuremath{3.7\times 10^{-9}}}
\newcommand{\ULNonGRTVSSOld}{\ensuremath{9.3\times 10^{-9}}}
\newcommand{\ULNonGRTVSTNew}{\ensuremath{2.7\times 10^{-9}}}
\newcommand{\ULNonGRTVSVNew}{\ensuremath{2.7\times 10^{-9}}}
\newcommand{\ULNonGRTVSSNew}{\ensuremath{6.9\times 10^{-9}}}
\newcommand{\ImprovementULNonGRTVST}{\ensuremath{1.5}}
\newcommand{\ImprovementULNonGRTVSV}{\ensuremath{1.4}}
\newcommand{\ImprovementULNonGRTVSS}{\ensuremath{1.3}}

\newcommand{\ULNonGRT}{\ensuremath{2.6\times 10^{-9}}}
\newcommand{\ULNonGRV}{\ensuremath{3.0\times 10^{-9}}}
\newcommand{\ULNonGRS}{\ensuremath{7.5\times 10^{-9}}}

\newcommand{\BayesSignalvsNoise}{\ensuremath{-0.947 \pm 0.004}}
\newcommand{\BayesNonGRvsGR}{\ensuremath{-0.498 \pm 0.010}}

\newcommand{\CalibrUncertMagLHOoFourB}{\ensuremath{4.7\%}}
\newcommand{\CalibrUncertMagLLOoFourB}{\ensuremath{4.0\%}}
\newcommand{\CalibrUncertPhaLHOoFourB}{\ensuremath{2.6^\circ}}
\newcommand{\CalibrUncertPhaLLOoFourB}{\ensuremath{2.9^\circ}}
\newcommand{\FinalCalibrUncertoFourB}{\ensuremath{6.2\%}}
\newcommand{\CalibrUncertMagLHOoFourC}{\ensuremath{13.3\%}}
\newcommand{\CalibrUncertMagLLOoFourC}{\ensuremath{1.9\%}}
\newcommand{\CalibrUncertPhaLHOoFourC}{\ensuremath{6.4^\circ}}
\newcommand{\CalibrUncertPhaLLOoFourC}{\ensuremath{1.2^\circ}}
\newcommand{\FinalCalibrUncertoFourC}{\ensuremath{13.4\%}}

\newcommand{\freqNinetyNineOfourbcAlzero}{\ensuremath{53.3}}
\newcommand{\ptestOfourbcAlzero}{\ensuremath{0.3}}
\newcommand{\sigOfourbcAlzero}{\ensuremath{4.9}}

\newcommand{\freqNinetyNineOfourbcAlTwoThirds}{\ensuremath{84.2}}
\newcommand{\ptestOfourbcAlTwoThirds}{\ensuremath{-0.1}}
\newcommand{\sigOfourbcAlTwoThirds}{\ensuremath{3.9}}

\newcommand{\freqNinetyNineOfourbcAlThree}{\ensuremath{347.9}}
\newcommand{\ptestOfourbcAlThree}{\ensuremath{-0.3}}
\newcommand{\sigOfourbcAlThree}{\ensuremath{0.7}}

\newcommand{\freqNinetyNineAllAlzero}{\ensuremath{56.2}}
\newcommand{\ptestAllAlzero}{\ensuremath{0.7}}
\newcommand{\sigAllAlzero}{\ensuremath{3.4}}

\newcommand{\freqNinetyNineAllAlTwoThirds}{\ensuremath{86.3}}
\newcommand{\ptestAllAlTwoThirds}{\ensuremath{0.2}}
\newcommand{\sigAllAlTwoThirds}{\ensuremath{2.6}}

\newcommand{\freqNinetyNineAllAlThree}{\ensuremath{338.2}}
\newcommand{\ptestAllAlThree}{\ensuremath{-0.3}}
\newcommand{\sigAllAlThree}{\ensuremath{0.5}}

\newcommand{\freqNinetyNineOoneOfouraAlzero}{\ensuremath{64.7}}
\newcommand{\ptestOoneOfouraAlzero}{\ensuremath{1.1}}
\newcommand{\sigOoneOfouraAlzero}{\ensuremath{4.8}}

\newcommand{\freqNinetyNineOoneOfouraAlTwoThirds}{\ensuremath{87.6}}
\newcommand{\ptestOoneOfouraAlTwoThirds}{\ensuremath{0.4}}
\newcommand{\sigOoneOfouraAlTwoThirds}{\ensuremath{3.6}}

\newcommand{\freqNinetyNineOoneOfouraAlThree}{\ensuremath{329.9}}
\newcommand{\ptestOoneOfouraAlThree}{\ensuremath{-0.3}}
\newcommand{\sigOoneOfouraAlThree}{\ensuremath{0.6}}

\newcommand{\ULUnifAllZero}{\ensuremath{7.4\times 10^{-9}}}
\newcommand{\ULUnifAllTwoThirds}{\ensuremath{5.5\times 10^{-9}}}
\newcommand{\ULUnifAllThree}{\ensuremath{7.5\times 10^{-10}}}
\newcommand{\ULUnifAllMarg}{\ensuremath{1.1\times 10^{-8}}}
\newcommand{\ULUnifOoneOFourAZero}{\ensuremath{1.1\times 10^{-8}}}
\newcommand{\ULUnifOoneOFourATwoThirds}{\ensuremath{7.9\times 10^{-9}}}
\newcommand{\ULUnifOoneOFourAThree}{\ensuremath{1.0\times 10^{-9}}}
\newcommand{\ULUnifOoneOFourAMarg}{\ensuremath{1.7\times 10^{-8}}}
\newcommand{\ImprovementUnifZero}{\ensuremath{1.5}}
\newcommand{\ImprovementUnifTwoThirds}{\ensuremath{1.4}}
\newcommand{\ImprovementUnifThree}{\ensuremath{1.3}}
\newcommand{\ImprovementUnifMarg}{\ensuremath{1.5}}

\newcommand{\ULLogUnifAllZero}{\ensuremath{2.8\times 10^{-9}}}
\newcommand{\ULLogUnifAllTwoThirds}{\ensuremath{2.0\times 10^{-9}}}
\newcommand{\ULLogUnifAllThree}{\ensuremath{2.6\times 10^{-10}}}
\newcommand{\ULLogUnifAllMarg}{\ensuremath{2.7\times 10^{-9}}}
\newcommand{\ULLogUnifOoneOFourAZero}{\ensuremath{3.9\times 10^{-9}}}
\newcommand{\ULLogUnifOoneOFourATwoThirds}{\ensuremath{2.8\times 10^{-9}}}
\newcommand{\ULLogUnifOoneOFourAThree}{\ensuremath{3.6\times 10^{-10}}}
\newcommand{\ULLogUnifOoneOFourAMarg}{\ensuremath{4.1\times 10^{-9}}}
\newcommand{\ImprovementLogUnifZero}{\ensuremath{1.4}}
\newcommand{\ImprovementLogUnifTwoThirds}{\ensuremath{1.4}}
\newcommand{\ImprovementLogUnifThree}{\ensuremath{1.4}}
\newcommand{\ImprovementLogUnifMarg}{\ensuremath{1.5}}

\newcommand{\bbhMedian}{\ensuremath 5.6}
\newcommand{\bbhUpperError}{\ensuremath 5.0}
\newcommand{\bbhLowerError}{\ensuremath 2.1}
\newcommand{\bbhPower}{\ensuremath -10}
\newcommand{\RBBHMedianOFourb}{\ensuremath 18.4}
\newcommand{\RBBHUpperErrorOFourb}{\ensuremath 5.1}
\newcommand{\RBBHLowerErrorOFourb}{\ensuremath 4.2}
\newcommand{\alphaMedianOFourb}{\ensuremath 2.7}
\newcommand{\alphaUpperErrorOFourb}{\ensuremath 0.9}
\newcommand{\alphaLowerErrorOFourb}{\ensuremath 0.7}

\newcommand{\bnsMedian}{\ensuremath 2.3}
\newcommand{\bnsUpperError}{\ensuremath 4.1}
\newcommand{\bnsLowerError}{\ensuremath 1.8}
\newcommand{\bnsPower}{\ensuremath -11}
\newcommand{\bnsrateMedian}{\ensuremath 38}
\newcommand{\bnsrateUpperError}{\ensuremath 69}
\newcommand{\bnsrateLowerError}{\ensuremath 30}

\newcommand{\nsbhMedian}{\ensuremath 3.6}
\newcommand{\nsbhUpperError}{\ensuremath 4.0}
\newcommand{\nsbhLowerError}{\ensuremath 2.3}
\newcommand{\nsbhPower}{\ensuremath -11}
\newcommand{\nsbhrateMedian}{\ensuremath 22}
\newcommand{\nsbhrateUpperError}{\ensuremath 24}
\newcommand{\nsbhrateLowerError}{\ensuremath 14}

\newcommand{\cbcMedian}{\ensuremath 6.3}
\newcommand{\cbcUpperError}{\ensuremath 5.0}
\newcommand{\cbcLowerError}{\ensuremath 2.2}
\newcommand{\cbcPower}{\ensuremath -10}
\newcommand{{\LnbfSchuNde}}{{\ensuremath -0.425}}
\newcommand{{\LnbfSchuNdeErr}}{{\ensuremath 0.007}}
\newcommand{{\LnbfSchuNdePLGaussAlpha}}{{\ensuremath -0.920}}
\newcommand{{\LnbfSchuNdePLGaussAlphaErr}}{{\ensuremath 0.010}}
\newcommand{{\LnbfSchuNdePLUnifAlpha}}{{\ensuremath -0.977}}
\newcommand{{\LnbfSchuNdePLUnifAlphaErr}}{{\ensuremath 0.010}}

\newcommand{{\LogTenbfSchuNde}}{{\ensuremath -0.185}}
\newcommand{{\LogTenbfSchuNdeErr}}{{\ensuremath 0.003}}
\newcommand{{\LogTenbfSchuNdePLGaussAlpha}}{{\ensuremath -0.400}}
\newcommand{{\LogTenbfSchuNdePLGaussAlphaErr}}{{\ensuremath 0.004}}

\begin{document}
%comment to remove the line numbers
%\linenumbers

\title{Updated Upper Limits on the Isotropic Gravitational-Wave Background from LIGO, Virgo, and KAGRA Data through April 2025}

\author{A.~G.~Abac\,\orcidlink{0000-0003-4786-2698}}
\affiliation{Max Planck Institute for Gravitational Physics (Albert Einstein Institute), D-14476 Potsdam, Germany}
\author{A.~Abe}
\affiliation{Department of Physics, Graduate School of Science, Osaka Metropolitan University, 3-3-138 Sugimoto-cho, Sumiyoshi-ku, Osaka City, Osaka 558-8585, Japan  }
\author{I.~Abouelfettouh}
\affiliation{LIGO Hanford Observatory, Richland, WA 99352, USA}
\author{F.~Acernese}
\affiliation{Dipartimento di Fisica ``E.R. Caianiello'', Universit\`a di Salerno, I-84084 Fisciano, Salerno, Italy}
\affiliation{INFN, Sezione di Napoli, I-80126 Napoli, Italy}
\author{K.~Ackley\,\orcidlink{0000-0002-8648-0767}}
\affiliation{University of Warwick, Coventry CV4 7AL, United Kingdom}
\author{A.~Adam}
\affiliation{OzGrav, University of Western Australia, Crawley, Western Australia 6009, Australia}
\author{C.~Adamcewicz\,\orcidlink{0000-0001-5525-6255}}
\affiliation{OzGrav, School of Physics \& Astronomy, Monash University, Clayton 3800, Victoria, Australia}
\author{S.~Adhicary\,\orcidlink{0009-0004-2101-5428}}
\affiliation{The Pennsylvania State University, University Park, PA 16802, USA}
\author{D.~Adhikari}
\affiliation{Max Planck Institute for Gravitational Physics (Albert Einstein Institute), D-30167 Hannover, Germany}
\affiliation{Leibniz Universit\"{a}t Hannover, D-30167 Hannover, Germany}
\author{R.~X.~Adhikari\,\orcidlink{0000-0002-5731-5076}}
\affiliation{LIGO Laboratory, California Institute of Technology, Pasadena, CA 91125, USA}
% ...

\collaboration{The LIGO Scientific Collaboration, the Virgo Collaboration, and the KAGRA Collaboration}

\begin{abstract}

We report results from a search for an isotropic stochastic gravitational-wave background using data collected by the LIGO--Virgo--KAGRA Collaboration. The analysis uses data from the first observing run through April 1, 2025, during the fourth observing run. New frequency-domain cuts are implemented to address a class of non-stationary spectral noise features that were not effectively identified and mitigated by existing data-quality checks in past analyses. Consequently, previously analyzed data from the fourth observing run are re-processed with the updated cuts. We find no evidence for a stochastic background signal and place upper limits on the gravitational-wave energy density. In particular, for a background following a power law with spectral index 2/3 as predicted by inspiralling compact binaries, we find $\Omega_\mathrm{GW}(25\,\mathrm{Hz}) \leq \ULLogUnifAllTwoThirds$, while scale-invariant backgrounds are constrained to $\Omega_\mathrm{GW}(25\,\mathrm{Hz}) \leq \ULLogUnifAllZero$, both at the 95\% credible level for a log-uniform prior on $\Omega_\mathrm{GW}$. Relative to the constraints from previous data recomputed with the new frequency-domain cuts, these limits improve by a factor of 1.4. We also update bounds on alternative gravity scenarios predicting non-standard polarization modes, and we verify that correlated magnetic noise sources remain below the sensitivity of this search. Combining these observational constraints with population models of compact binary coalescences informed by the latest gravitational-wave transient catalog, GWTC-5.0, we predict the amplitude of the compact binary background to be $\Omega_\mathrm{CBC}(25\,\mathrm{Hz}) = \cbcMedian^{+\cbcUpperError}_{-\cbcLowerError} \times 10^{\cbcPower}$ at the 90\% credible level.

\end{abstract}

\maketitle

%%%%%%%%%%%%%%%%%%%%%%%%%%%%%%%%%%%%%%%%%%%%%%%
\section{Introduction}
\label{sec:introduction}
The gravitational-wave background (GWB) is a random superposition of a large number of weak, independent, and unresolved gravitational-wave (GW) signals, of both astrophysical and cosmological origin. Potential astrophysical contributors to the GWB include unresolved compact binary mergers (CBCs)~\cite{PhysRevD.84.124037,Zhu:2011,PhysRevD.84.084004, PhysRevD.84.124037, PhysRevD.85.104024,Zhu_CBC_2}, core-collapse supernovae~\cite{PhysRevD.72.084001,PhysRevD.73.104024,PhysRevD.73.104024, PhysRevD.84.124037, Zhu_CBC_2, Tania_CC_2, Tania_CC}, rotating neutron stars~\cite{8201910,PhysRevD.72.084001, PhysRevD.73.104024, Zhu:2011pt, PhysRevD.86.104007,PhysRevD.87.063004}, and other exotic sources such as superradiance of axion clouds around black holes~\cite{PhysRevD.83.044026,Brito:2015oca,PhysRevLett.117.061101,PhysRevLett.119.131101,Takahashi:2024fyq}. Possible cosmological sources include cosmic strings~\cite{Kibble_1976,Vachaspati:1985,Hindmarsh:1994re,PhysRevLett.85.3761, Damour:2004kw,Siemens:2007,Olmez:2010,Kawasaki:2010,Vachaspati:2015cma,Blanco-Pillado:2017,PhysRevD.97.102002,Auclair:2019wcv,PhysRevLett.126.241102,Schmitz_CS}, first-order phase transitions in the early Universe~\cite{PhysRevD.30.272, Hogan:1986qda, PhysRevLett.65.3080, PhysRevD.45.4514, PhysRevD.49.2837,Mazumdar:2018dfl,Hindmarsh:2020hop,PhysRevLett.126.151301, PhysRevD.107.023511}, primordial black hole binaries~\cite{1975ApJ...201....1C,Raidal:2017mfl, Carr:2020gox,zphk-3ld9}, primordial scalar perturbations~\cite{Matarrese:1997ay, Baumann:2007zm,Kohri:2018awv,PhysRevLett.128.051301, SIGWB_Ryoto, Mukherjee:2021ags}, epochs with a stiff equation of state~\cite{Giovannini:1998bp,Giovannini:1999bh,Figueroa:2018twl,Figueroa:2019paj,PhysRevD.110.103503}, and inflation~\cite{Starobinsky:1979ty,Mukhanov:1981xt, ABBOTT1984541,Barnaby:2011qe,Barnaby:2011vw,Maleknejad:2016qjz,Bartolo:2016ami,Thorne:2017jft,PhysRevD.110.084063}. 

A detection of the GWB could provide a unique observational window onto the Universe at energy scales and epochs that are inaccessible to electromagnetic observations or terrestrial particle accelerators~\cite{Maggiore:1999vm}. In particular, cosmological backgrounds would carry direct information about the physical processes operating in the very early Universe. At the same time, an astrophysical GWB probes the population properties of compact objects at redshifts beyond the reach of individually resolvable GW events~\cite{Zhu:2011,Mukherjee:2019oma,PhysRevLett.125.141102, Sah:2023bgr,Sah:2025agw}.
The Laser Interferometer Gravitational-wave Observatory (LIGO)~\cite{Aasi_2015}, Virgo~\cite{Acernese_2015}, and KAGRA~\cite{KAGRA:2020tym} collaborations (LVK Collaboration~\cite{ProspectsLVK:2016}) have previously placed upper limits on isotropic~\cite{O4a_iso_paper} and anisotropic~\cite{wgp7-t8r8} GWBs as well as on various cosmological models~\cite{CosmoImplication} using the data from the first three observing runs (O1, O2, and O3) and the first part of the fourth observing run (O4a).

Given the large number of incoherent and unresolved sources contributing to the GWB, the signal is naturally described within a statistical framework~\cite{PhysRevD.59.102001, Detection_methods_Joe_Neil, PhysRevD.92.042001}. Under the assumption that the individual contributions are independent, the central limit theorem implies that the GWB can be modeled as a Gaussian stochastic process~\cite{Detection_methods_Joe_Neil} whose strain has zero mean~\cite{Jenkins2022CosmologyAF}, which we further assume to be stationary, isotropic, and unpolarized. Its statistical properties are fully characterized by the two-point correlation function, or equivalently by its energy density spectrum. The optimal estimator is therefore obtained via a cross-correlation analysis between pairs of spatially separated detectors~\cite{PhysRevD.59.102001, Detection_methods_Joe_Neil, PhysRevD.92.042001}, exploiting the fact that a stochastic background induces correlated signals across the network. A summary of this formalism is provided in Sec.~II of Ref.~\cite{O4a_iso_paper}, with comprehensive reviews available in Refs.~\cite{PhysRevD.59.102001, Detection_methods_Joe_Neil, PhysRevD.92.042001, Zhong:2026owg}.

In this work, we extend previous isotropic GWB searches~\cite{O4a_iso_paper} by incorporating data from the second and early third parts of the LVK Collaboration's fourth observing run (hereafter denoted O4b~\cite{LIGOScientific:2026jgl} and O4c1, respectively). We restrict the O4c data set to the period before the scheduled O4c observing break, during which operations were paused from April 1 to June 4, 2025 for maintenance and commissioning work, including the repair of a faulty beam-tube gate valve at LIGO Livingston and other detector improvements intended to reduce risks in the transition from O4 to O5~\cite{LVK_O4c_schedule_update}.

We perform a hybrid frequentist-Bayesian cross-correlation analysis~\cite{Detection_methods_Joe_Neil, Sufficient_statistic_Andrew, O4a_iso_paper} implemented in the open-source analysis pipeline \texttt{pygwb}~\cite{Renzini2024, Renzini:2023qtj}, and update upper limits on the GWB energy density. Our analysis assumes that detector noise is uncorrelated with the GW signal and between different detectors. While the latter assumption is standard in GWB searches~\cite{O4a_iso_paper}, we explicitly assess its validity in this work and verify that correlated magnetic noise~\cite{Thrane_2013, Thrane_2014, Janssens:2022tdj, Janssens:2025uxm, Coughlin:2016vor, PhysRevD.97.102007, Janssens:2021cta} remains negligible. 

General relativity (GR) predicts only tensor polarizations, but many alternative theories of gravity allow for additional scalar and vector modes. We therefore extend our isotropic GWB signal model to include tensor, vector, and scalar polarizations and place corresponding upper limits on the amplitude of each component~\cite{PhysRevD.79.082002,Callister:2017ocg}.

In the most sensitive frequency band of the current ground-based detector network, the GWB is expected to be dominated by the superposition of unresolved CBCs~\cite{regimbau:hal-01084645}. 
Building on the methodology of Ref.~\cite{O4a_iso_paper}, we update predictions for the spectrum from binary black holes (BBHs), binary neutron stars (BNSs), and neutron star--black hole (NSBH) coalescences based on the astrophysical distribution of compact binary mergers reported in the recent GWTC-5.0 catalog~\cite{GWTC-5-Introduction:2026,GWTC-5-Population:2026}. 
We compare such estimates with the latest projections of sensitivity curves for future observing runs~\cite{O5strain-projection} and revise previous statements about the detection prospects.

This paper is organized as follows. In Sec.~\ref{sec:data_quality}, we describe the data quality of the additional data included in our analysis. Sec.~\ref{sec:results} presents the main results of our search, including the cross-correlation spectra (Sec.~\ref{sec:frequentist_analysis}), updated upper limits on the GWB (Sec.~\ref{sec:bayesian_analysis}), improved constraints on additional polarization components (Sec.~\ref{sec:non_GR}), and a joint analysis of magnetic noise and stochastic signals (Sec.~\ref{sec:joint_mag_stoch}). Finally, in Sec.~\ref{sec:astro_implications}, we update the predicted amplitude of the compact binary background.

%%%%%%%%%%%%%%%%%%%%%%%%%%%%%%%%%%%%
\section{Data quality}
\label{sec:data_quality}
In this section, we describe the data quality of the strain data from the LIGO Hanford and LIGO Livingston detectors during O4b and O4c1. These data sets are combined with those from previous observing runs, namely O1--O4a LIGO data together with Virgo O3 data~\cite{PhysRevLett.118.121101, PhysRevD.100.061101, PhysRevD.104.022004, O4a_iso_paper}. The O4b observing period started at 17:00 UTC on April 10, 2024 and ended at 17:00 UTC on January 28, 2025. Throughout O4b there were two notable breaks during which the interferometers were taken out of observing mode to diagnose and address issues impacting the sensitivity. At LIGO Hanford, operations were interrupted between July 13 and August 24, 2024 due to issues related to the output Faraday isolator. At LIGO Livingston, a separate break occurred between October 11 and October 29, 2024, associated with a fast shutter malfunction. The O4c1 observing period ran without interruption after the end of O4b from 17:00 UTC on January 28, 2025, to 15:00 UTC April 1, 2025. Further details on detector performance and commissioning activities during O4b and O4c1 can be found in Ref.~\cite{Glanzer:2026fhy}.  In the following, we shall refer to the LIGO Hanford detector as LHO and the LIGO Livingston detector as LLO. 

The Virgo detector also collected data during O4b and O4c1. However, in the remainder of this work we use only the data from the LHO and LLO detectors during that period. The decision not to include the Virgo data in our analysis is motivated by its marginal contribution to improving the search sensitivity. Quantitatively, we estimate that including Virgo data would lead to improvements in $\sigma_o$ (Eq. (8) in Ref.~\cite{O4a_iso_paper}) of only 1\% during O4b. Virgo data from the third observing run are nevertheless incorporated in our combined analysis. In addition, KAGRA data were not included in this study, as their contribution would be even smaller than that of the Virgo data.

\subsection{Data}
The total coincident observing time for the Hanford–Livingston detector baseline from the two periods was 140.99 days, before applying data-quality cuts. For this analysis, we use the calibrated strain data product~\cite{LIGOScientific:2026jgl} with low-latency subtractions of sinusoidal calibration lines, 60 Hz power lines~\cite{Vajente_2020}, and residual noise coupling where available~\cite{Capote_2025}. The data channel is \texttt{GDS-CALIB\_STRAIN\_CLEAN\_AR}, and the search is restricted to the frequency range $20\text{--}1726\,\mathrm{Hz}$.

\subsection{Time-domain cuts}
We apply a series of time-domain data-quality cuts at different stages of the analysis, following the procedures described in Ref.~\cite{O4a_iso_paper}. We summarize in Table~\ref{tab:data_cuts} the amount of data remaining after the application of the CAT1 vetoes, which remove data collected while the detectors were operating outside nominal conditions~\cite{Glanzer:2026fhy}, and the final livetime after data preprocessing, including gating, which suppresses periods affected by excess noise by applying an inverse Tukey window to the affected regions of the strain data, effectively setting the corresponding data to zero. The postprocessed data are then subjected to the $\Delta\sigma$ cut, a non-stationarity cut that removes periods during which the square root of the variance varies by more than 20\% between neighbouring segments. We highlight that 0.078\% of LHO data and 0.068\% of LLO data were gated and that 9\% of the analyzed segments were discarded following the delta-sigma cut for broadband non-stationarities. These noise artifacts produce excess power distributed over multiple frequency bins, rather than being confined to a single frequency bin of width 1/32 Hz used in the analysis. This leaves a total of 117.05 days of usable data for the analyses presented in this work.

\begin{table}[h]
\centering
\begin{tabular}{lc}
\hline\hline
\textbf{Selection Step} & \textbf{ Days} \\
\hline
Coincident observing time &  140.99  \\
Time after CAT1 vetoes &  138.37\\
Final livetime &  117.05 \\
\hline\hline
\end{tabular}
\caption{Total duration of coincident data between detectors during O4b and O4c1, together with the amount of data remaining after the application of time-domain data-quality cuts.}
\label{tab:data_cuts}
\end{table}

\subsection{Frequency-domain cuts}

We perform a targeted re-assessment of the frequency domain cuts applied in Ref.~\cite{O4a_iso_paper} to the analysis band 20--1726\,Hz and update them for this work. The procedure is motivated by features observed in the cross-spectral density (CSD) and in the individual detector power spectral densities (PSDs), as well as by their impact on the posterior distributions of the parameters of the model adopted to describe the GWB. Specifically, we assume a power-law model of the form
\begin{equation}
\Omega_{\rm GW}(f) = \Omega_{\rm ref} \left( \frac{f}{25\,\mathrm{Hz}} \right)^{\alpha},
\label{power_law}
\end{equation}
where $\Omega_{\rm ref}$ is the GWB amplitude at the reference frequency of $25\,\mathrm{Hz}$, and $\alpha$ is the spectral index.

In addition to dedicated investigations of specific spectral features, we implement a procedure to identify frequency bands exhibiting non-stationary behavior throughout the observing period. This approach is based on the temporal variability of the PSD and provides a complementary method to construct our frequency-domain cuts. For each detector, we compute the standard deviation of the PSD normalized by its median value across time segments produced by \texttt{pygwb}~\cite{Renzini:2023qtj, Renzini2024},
\begin{equation}
\mathrm{StdRatio}(f)
=
\mathrm{std}\left(\frac{\mathrm{PSD}(f)}{\mathrm{median}(\mathrm{PSD}(f))}\right).
\end{equation}
The median and standard deviation are computed over all time segments for a given frequency $f$. We then estimate a smooth local baseline for $\mathrm{StdRatio}(f)$ by identifying broadband, peak-like structures - regions that rise well above a running median-filtered baseline - and interpolating linearly across them. This interpolation is used only to construct a baseline representative of the typical variability level, and no features are flagged for removal at this point. The original $\mathrm{StdRatio}(f)$ spectrum is divided by this smooth baseline to form a normalized variability spectrum, in which frequencies with excess temporal variability relative to nearby frequencies are enhanced. An empirical threshold of 1.5 is then applied to flag outlier bins, which are subsequently merged into contiguous frequency bands to produce an initial, unvetted list. Each flagged band is vetted by cross-checking against the catalog of known instrumental spectral artifacts~\cite{narrow_band_artifacts_Goetz:2026aad}. This resulting list, combined with the targeted studies described below, forms the \textit{strict} notch list, which is in turn combined with the baseline notch list of Ref.~\cite{O4a_iso_paper} to produce the final set of frequency-domain cuts applied in this analysis. This procedure provides a complementary view of detector performance, highlighting frequency bands that fluctuate significantly over the course of the observing run, or sidebands of known artifacts such as the power-line harmonics, that are not always apparent in the run-averaged PSDs or coherence spectra used to construct the baseline notch list of Ref.~\cite{O4a_iso_paper}.

\paragraph{Updated power-line harmonics ($20\text{--}1726\,\mathrm{Hz}$).} The notches around the $60\,\mathrm{Hz}$ power-line harmonics are re-evaluated across the full analysis band. In O4 data, several harmonics exhibited slow frequency wandering and sideband structure, leading to residual contamination outside the nominal narrow notches. These features are consistently identified by the strict procedure described above as regions of excess variability, which support the need for widened notches beyond the nominal harmonic frequencies.

\paragraph{Low-frequency region ($<100\,\mathrm{Hz}$).}
Below $100\,\mathrm{Hz}$, which contributes most significantly to the sensitivity for physically motivated spectral indices $\alpha = 0$ and $\alpha = 2/3$, we identified prominent features in the CSD between the LHO and LLO strain data. These include mechanical resonances of optical components as well as intermodulation products of low-frequency calibration lines. These structures are also captured by the strict data-quality procedure described above. 

\paragraph{High-frequency region ($400-900\,\mathrm{Hz}$).} Several instrumental features in this band were found to drive excess power in steep spectral indices such as $\alpha = 5$. The main features in this band were associated with suspension violin modes and a time-dependent excess in the Livingston detector:
\begin{itemize}
    \item Excess narrowband structure was observed around the fundamental suspension violin modes near $500\,\mathrm{Hz}$. These modes are mechanical resonances of the quadruple suspension fibers, with higher-order harmonics appearing around $1000\,\mathrm{Hz}$ and above \cite{PhysRevLett.116.131103}. Throughout O4, during periods when the violin modes were excited, additional narrow spectral features appeared in nearby frequency bands. Narrowband spectral artifact studies found that many of these features were consistent with intermodulation involving violin modes, low-frequency calibration lines, and their harmonics~\cite{narrow_band_artifacts_Goetz:2026aad}.

    \item A broadband, time-dependent excess in LIGO Livingston between $540$--$640\,\mathrm{Hz}$. Although auxiliary-channel studies showed modest correlations with the temperature monitoring channel at the corner stations, the exact source and coupling mechanism remain unidentified~\cite{Glanzer:2026fhy}. 
\end{itemize}
The strict notching procedure independently identifies these regions as exhibiting high temporal variability. This methodology produces two distinct final notch lists: one for the O4a data set, differing from that reported in Ref.~\cite{O4a_iso_paper}, and one for the O4b--O4c1 data set. Accordingly, we also report updated O1--O4a results in this paper that account for the stricter notch list. In total, we remove 22.2\% of the $20\text{--}1726\,\mathrm{Hz}$ analysis band for O4a and 22.8\% for O4b--O4c1, of which the strict notching procedure contributes an additional 2.1\% and 4.5\%, respectively.

\subsection{Magnetic noise budget}
\label{sec:magnetic_budget}
As mentioned above, since searches for a GWB rely on correlations of strain data between spatially separated detectors, the analysis is susceptible to bias from correlated noise sources \cite{Thrane_2014}. A well-known example is magnetic noise associated with Schumann resonances--global standing electromagnetic waves in the cavity formed by Earth’s surface and the ionosphere, excited primarily by lightning activity \cite{1952ZNatA...7..250S, 1952ZNatA...7..149S, PhysRevD.97.102007}. Owing to their global nature, these resonances can induce correlated signals even between widely separated detectors \cite{Thrane_2013}. 

Environmental magnetic fields can couple to the detector response through several mechanisms~\cite{envsnoise}, including pickup in cables and other conductive components in the detector building, direct forces on the actuation magnets of the test-mass mirrors (as in Virgo~\cite{ENV_noise}), and forces on magnets or magnetically susceptible material higher up in the test-mass suspension system (as in LIGO \cite{Thrane_2013}). The strength of this coupling is quantified through coupling functions, namely the outside-to-inside and inside-to-strain transfer functions, capturing the couplings of magnetic field from outside the experimental buildings to inside the buildings and from inside the buildings to the strain detector channel, respectively. A detailed description of these quantities can be found in Sec.~III~F of Ref.~\cite{O4a_iso_paper}. For the outside-to-inside coupling functions, we adopt the same values and uncertainties used in the O4a and O3 magnetic noise budget analyses \cite{PhysRevD.104.022004, O4a_iso_paper}. For the inside-to-strain coupling, new values have been measured for each interferometer using the procedure described in Refs. \cite{Janssens:2022tdj, Nguyen_2021, ENV_noise}.

The magnetometer calibration factors were remeasured during O4b, yielding values with improved accuracy (approximately $5\%$ uncertainty) relative to those used in the O4a and O3 analyses, which had been derived prior to O3. The updated O4b measurements give calibration factors smaller than those from O3, corresponding to O3-to-O4 ratios of 1.26 for LHO in the magnetometer along the X arm and 1.24 along the Y arm. For LLO, the corresponding ratios are 4.07 and 3.24. This reduction results in a lower overall magnetic noise budget compared to the estimate for O4a.

\begin{figure}[h]
    \centering
    \includegraphics[width =1\linewidth]{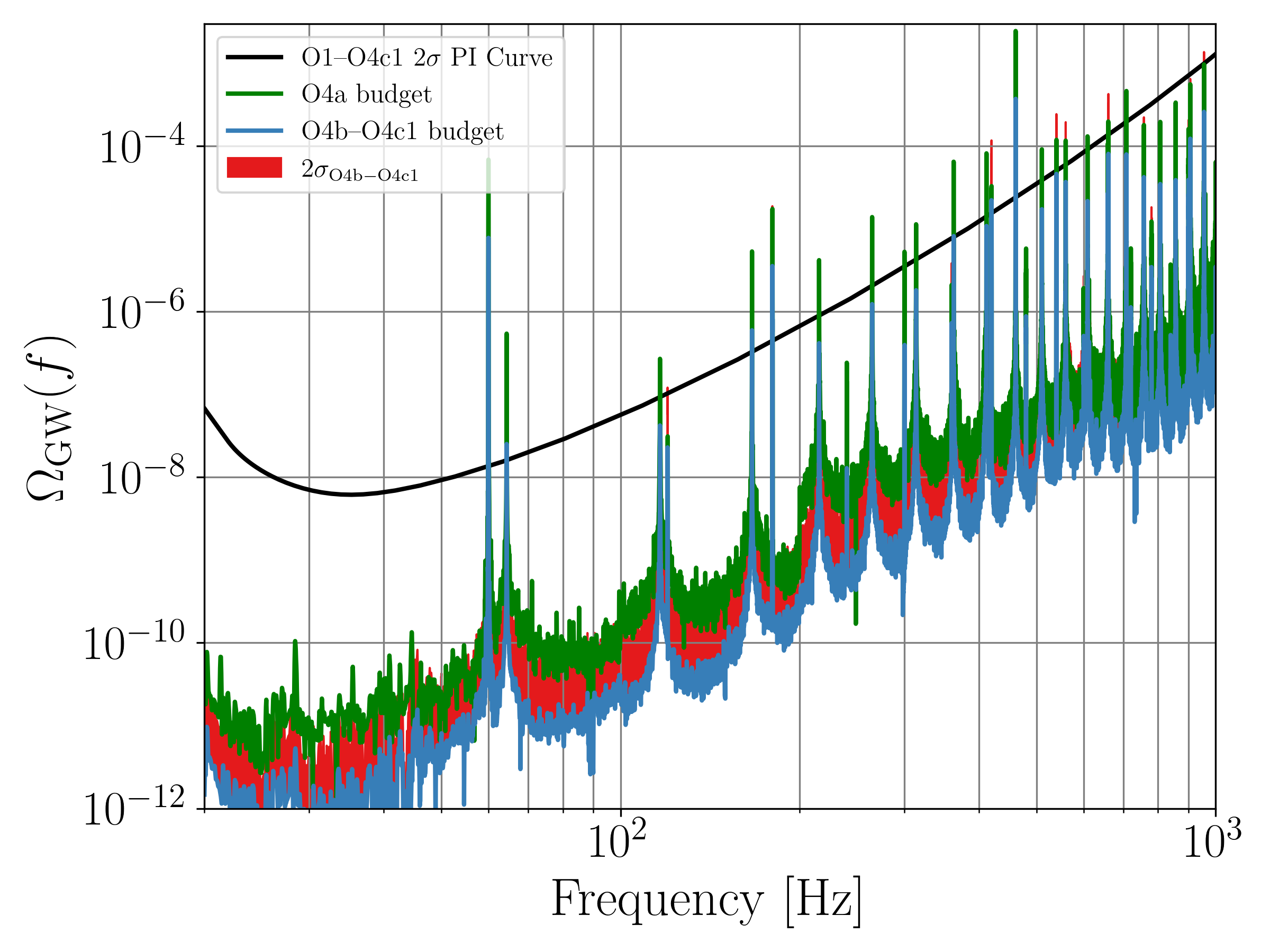}
    \caption{Comparison of the magnetic noise budget calculated for O4a (green curve) and O4b--O4c1 (blue curve), which shows a difference in magnitude due to the remeasuring of the magnetometer calibration factors. The red band represents the $2\sigma$ uncertainty (shown only upwards) of the O4b--O4c1 budget.}
    \label{fig:magnetic_a_vs_bc}
\end{figure}

The computation of the magnetic noise budget is described in detail in Sec.~III~F of Ref.~\cite{O4a_iso_paper}, and the same approach is adopted in this work. Uncertainty estimation follows an identical procedure: the outside-to-inside coupling uncertainty measured during O3 at LLO is used, since equivalent measurements were not available for LHO or during O4. This is combined with a conservative intrinsic uncertainty factor of 2 to account for the mismatch between the location of the magnetometers in the building (used for the measurements) and the location where the magnetic field couples to the strain. The final magnetic noise budget is shown in Fig.~\ref{fig:magnetic_a_vs_bc}. The calculated budget (blue solid curve) is compared with the $2\sigma$ power-law integrated (PI) sensitivity curve~\cite{thrane_sensitivity_2013}, obtained over the full O1–O4c1 observing period (black solid curve), which quantifies the sensitivity of the search to power law (PL) GWB. We also show the $2\sigma$ uncertainty on the O4b--O4c1 magnetic noise budget (red band). For comparison, the O4a magnetic noise budget is shown as a green solid line.

\subsection{Calibration uncertainties}
Strain calibration uncertainties arise from both statistical fluctuations and systematic errors in the LVK detector strain calibration~\cite{Sun:2020wke, Sun:2021qcg}.

In our Bayesian upper limits estimation, these uncertainties are incorporated by introducing a positive-definite calibration factor $\Lambda$ in the likelihood, as described in Sec.~III~G of Ref.~\cite{O4a_iso_paper} (see also Refs.~\cite{PhysRevD.103.063016,Renzini:2023qtj} for additional details). The parameter $\Lambda$ is marginalized over using the prior specified in Sec.~III~G of Ref.~\cite{O4a_iso_paper}, with the corresponding calibration variance, $\sigma_\Lambda$, determined as described in Ref.~\cite{LIGOScientific:2026jgl}. Note that the calibration uncertainty modifies the variance of our estimator, as described in Eq.~(5) of Ref.~\cite{Whelan:2012ur}. We then use this modified variance for our estimator in the frequentist approach.

\begin{table}[t]
\centering
\begin{tabular}{c|c|c|c|c}
 & \multicolumn{2}{c|}{\textbf{O4b}} & \multicolumn{2}{c}{\textbf{O4c1}} \\
\textbf{Interferometer} & \textbf{LHO} & \textbf{LLO} & \textbf{LHO} & \textbf{LLO} \\
\hline \hline
\textbf{Magnitude} 
& \CalibrUncertMagLHOoFourB & \CalibrUncertMagLLOoFourB 
& \CalibrUncertMagLHOoFourC & \CalibrUncertMagLLOoFourC \\
\hline
\textbf{Phase} 
& \CalibrUncertPhaLHOoFourB & \CalibrUncertPhaLLOoFourB 
& \CalibrUncertPhaLHOoFourC & \CalibrUncertPhaLLOoFourC \\
\end{tabular}
\caption{Maximum strain calibration uncertainties at 68\% credible level (CL) for the magnitude and phase of detector strain for O4b and O4c1.}
\label{tab:calib_uncertainties}
\end{table}

Table~\ref{tab:calib_uncertainties} summarizes the maximum uncertainties for the LHO and LLO detectors during O4b and O4c1. We note that the magnitude uncertainty at LHO during O4c1 is larger than usual. This is because, in the first part of O4c, high-frequency measurements at H1 were not available for certain time periods, resulting in reduced coverage above 1\,kHz and correspondingly larger uncertainties in that frequency band. These uncertainties were later reduced in the second part of the third period of O4 (O4c2) once additional high-frequency measurements were included. While the overall quality of the O4c1 data is comparable to that of O4b, this effect is partly related to the way the data are analyzed in discrete detector epochs, within which the available calibration measurements vary depending on data availability at each stage. Importantly, we have verified that this increased uncertainty does not substantially affect our final Bayesian results.

As for the phase uncertainties, which remain below or close to $\lesssim 5^\circ$, they do not introduce a significant bias in GWB analyses~\cite{Yousuf:2023nmz}. Therefore, our focus is on the magnitude component. By combining the magnitude uncertainties from each detector in quadrature, we obtain an overall calibration variance of $\sigma_\Lambda = \FinalCalibrUncertoFourB$  during O4b and $\sigma_\Lambda = \FinalCalibrUncertoFourC$ during O4c1.

%%%%%%%%%%%%%%%%%%%%%%%%%%%
\section{Results}
\label{sec:results}
We present the results obtained using the methodology described in Ref.~\cite{O4a_iso_paper}, applied to data from the O4b and O4c1 observing periods in combination with data from the O1--O4a runs. No evidence for a GWB signal is found. Consequently, we set Bayesian upper limits on a range of models. 

\subsection{Frequentist results}
\label{sec:frequentist_analysis}
We report in Table~\ref{tab:Frequentist_results} the values of the broadband estimators $\hat{\Omega}_{\rm ref}$~\cite{O4a_iso_paper} and their standard deviations for the O4b--O4c1 data set and the O1--O4c1 data set, evaluated at the $25\,\mathrm{Hz}$ reference frequency. We adopt three values of the spectral index (see Eq.~(\ref{power_law})): $\alpha = 0$, corresponding to a scale-invariant signal such as a cosmological GWB from slow-roll inflation or cosmic strings~\cite{Caprini-Figueroa:2018mtu}; $\alpha = 2/3$, consistent with an astrophysical GWB from compact binary coalescences (CBCs)~\cite{regimbau:hal-01084645}; and $\alpha = 3$, corresponding to a flat strain power spectrum~\cite{PhysRevD.59.102001}.

For each spectral index, we also report the frequency $f_{99\%}$, defining the frequency range $[20, f_{99\%}]\,\mathrm{Hz}$ within the $[20, 1726]\,\mathrm{Hz}$ band that contributes to 99\% of the search sensitivity.

\begin{figure*}[t]
    \centering 
    \includegraphics[width=0.95\textwidth]{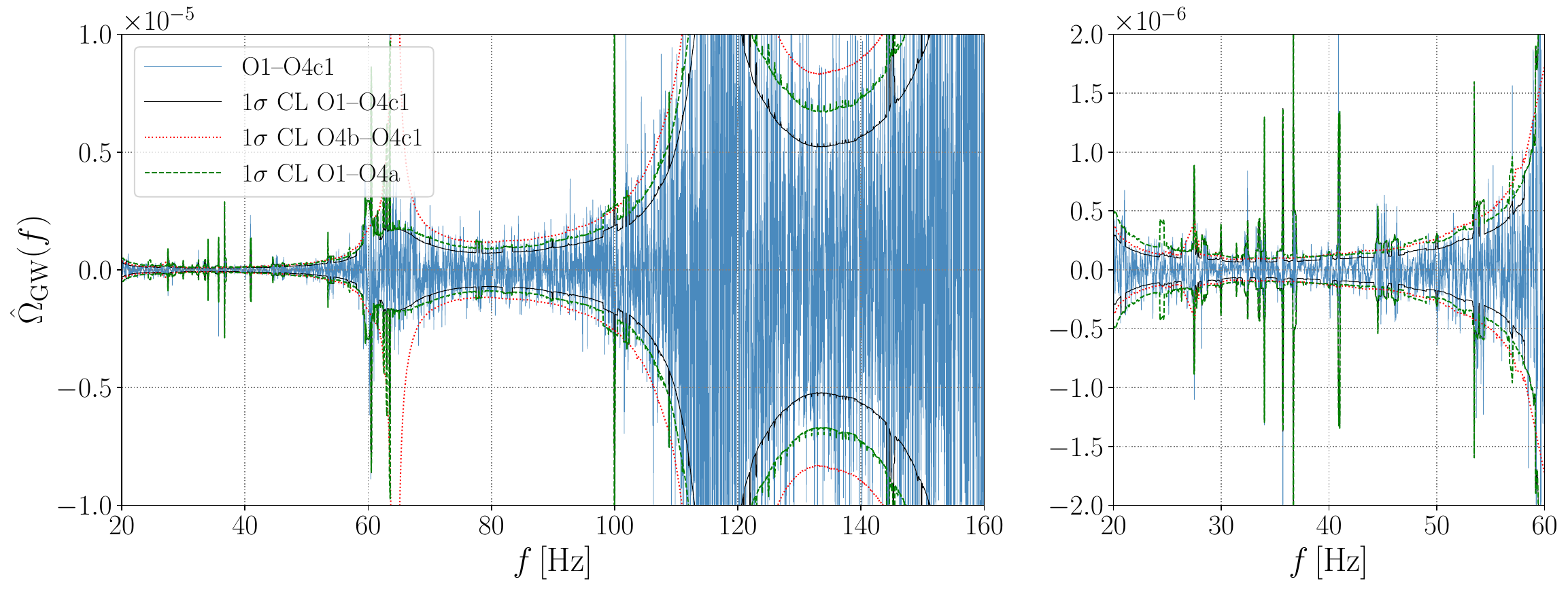} 
    \caption{Cross-correlation spectrum obtained from the combined O1--O4c1 data. The black solid lines denote the boundaries of the O1--O4c1 $1\sigma$ region. The red dotted lines indicate the corresponding $1\sigma$ region for the O4b--O4c1 data alone, while the green dashed lines show the updated $1\sigma$ region for the O1--O4a data, reflecting the use of the stricter notch list relative to Ref.~\cite{O4a_iso_paper}, as described in Sec.~\ref{sec:data_quality}. The spectrum (blue curve) remains consistent with zero across all frequency bins, showing no evidence for a GWB signal. The left panel shows the spectrum over the 20--160\,Hz range, while the right panel focuses on frequencies below 60\,Hz to emphasize the contribution from the O4b--O4c1 data in this band.} 
    \label{fig:cross_corr_spec} 
\end{figure*}

The point estimates $\hat{\Omega}_{\rm ref}$ are consistent with zero within $1\sigma$ uncertainties for both the O4b--O4c1 data set and the combined O1--O4c1 data set, with the results for $\alpha = 0$ and $\alpha = 2/3$ exhibiting a marginally positive signal-to-noise ratio (SNR). We also show the cross-correlation spectrum from O1--O4c1 in $\Omega_{\rm GW}$ units in Fig.~\ref{fig:cross_corr_spec}, together with the $1\sigma$ confidence regions from O4b--O4c1 and O1--O4c1. The cross-correlation spectrum and the $1\sigma$ confidence regions are consistent with zero, as expected for data dominated by uncorrelated Gaussian noise.

We also report updated results for the O1--O4a data set in Table~\ref{tab:Frequentist_results} and Fig.~\ref{fig:cross_corr_spec}, as we adopt a new (stricter) O4a notch list, as described in Sec.~\ref{sec:data_quality}. We find that the updated notch list has a mild impact on the uncertainties of the estimators. The corresponding values of $f_{99\%}^{\mathrm{O1\text{--}O4a}}$ also differ from those reported in Ref.~\cite{O4a_iso_paper}.

\begin{table*}
    \centering
    \begin{tabular}{c|c|c|c|c|c|c}
        ~$\alpha$~ & ~ $ f_{99\%} ^ {\mathrm{O1-O4a}}$\,[Hz] ~ & ~ $\hat{\Omega}^{\rm  O1-O4a}/10^{-9}$ & ~ $ f_{99\%} ^ {\mathrm{O4b-O4c1}}$\,[Hz] ~ & ~ $\hat{\Omega}^{\rm  O4b-O4c1}/10^{-9}$ ~ & ~ $ f_{99\%} ^ {\mathrm{O1-O4c1}}$\,[Hz] ~ & ~ $ \hat{\Omega}^{\rm O1-O4c1}/10^{-9}$~ \\
        \hline
        \hline
        0 & \freqNinetyNineOoneOfouraAlzero & \ptestOoneOfouraAlzero$\pm$\sigOoneOfouraAlzero & \freqNinetyNineOfourbcAlzero & \ptestOfourbcAlzero$\pm$\sigOfourbcAlzero & \freqNinetyNineAllAlzero & \ptestAllAlzero$\pm$\sigAllAlzero \\
2/3 & \freqNinetyNineOoneOfouraAlTwoThirds & \ptestOoneOfouraAlTwoThirds$\pm$\sigOoneOfouraAlTwoThirds & \freqNinetyNineOfourbcAlTwoThirds & \ptestOfourbcAlTwoThirds$\pm$\sigOfourbcAlTwoThirds & \freqNinetyNineAllAlTwoThirds & \ptestAllAlTwoThirds$\pm$\sigAllAlTwoThirds \\
3 & \freqNinetyNineOoneOfouraAlThree & \ptestOoneOfouraAlThree$\pm$\sigOoneOfouraAlThree & \freqNinetyNineOfourbcAlThree & \ptestOfourbcAlThree$\pm$\sigOfourbcAlThree & \freqNinetyNineAllAlThree & \ptestAllAlThree$\pm$\sigAllAlThree \\
    \end{tabular}
    \caption{Results from the isotropic search obtained with the standard cross-correlation optimal filter analysis, assuming PL spectra with indices $\alpha = \{0,2/3,3\}$. For each spectral index, we report the point estimate of the amplitude $\hat{\Omega}_{\rm ref}$ at 25\,Hz together with its associated $1\sigma$ uncertainty. We also provide the frequency interval spanning from 20\,Hz up to $f_{99\%}$, which contains 99\% of the total sensitivity. The estimates are constructed by combining statistically independent frequency-bin measurements according to Eqs.~(10) and (11) of Ref.~\cite{O4a_iso_paper}. The final two columns present the results obtained from the post-processing recombination of all observing runs (O1–O4c1), following the methodology of Ref.~\cite{PhysRevD.104.022004}. We note that Virgo data contribute to the O3 measurements, leading to three independent baseline contributions in that observing period. We have also included the results from the combination of data from O1 to O4a, an update of the results presented in Ref.~\cite{O4a_iso_paper}.
    }
    \label{tab:Frequentist_results}
\end{table*}

\subsection{Bayesian results}
\label{sec:bayesian_analysis}
\begin{figure}
    \centering
    \includegraphics[width=\columnwidth]{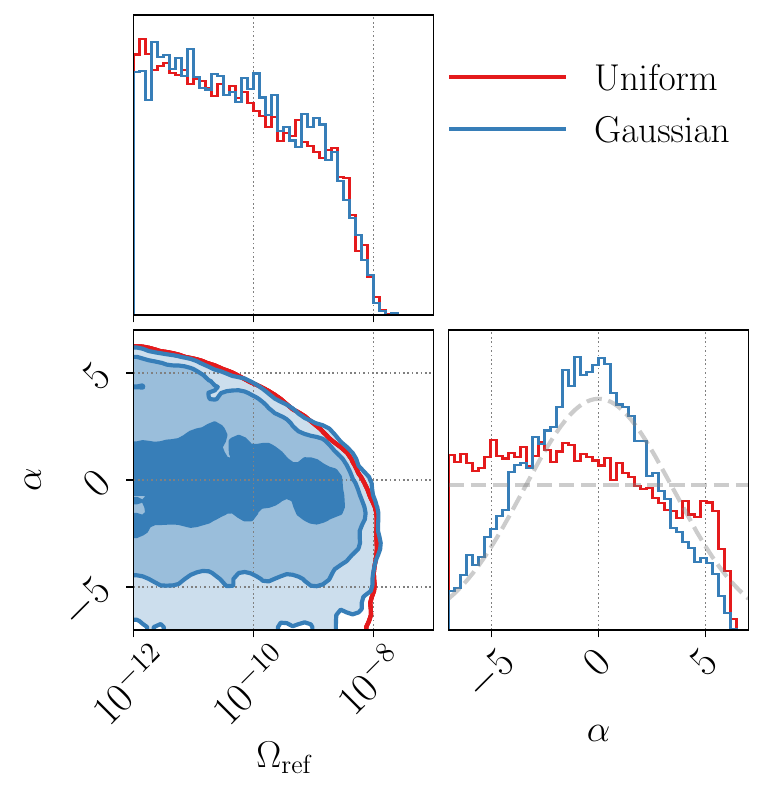}
    \caption{Posterior distributions on the PL model parameters \{$\Omega_{\rm ref}$, $\alpha$\} describing a GWB under the assumptions of a log-uniform prior on
$\Omega_{\rm ref}$ and either a Gaussian (blue curve) or a uniform (red curve) prior on $\alpha$, with the $\alpha$ priors in the corresponding one-dimensional histogram (gray dashed lines). The blue shaded regions indicate the 1, 2, and 3$\sigma$ credible intervals, from darkest to lightest. }
    \label{fig:PE_contour_plots}
\end{figure}

The data show no evidence for a signal; therefore, we set Bayesian upper limits at the 95\% credible level (CL) based on the frequentist estimator result. These constraints are summarized in Table~\ref{tab:upper_limits_Omega_ref}, and the corresponding posterior distributions are shown in Fig.~\ref{fig:PE_contour_plots}. The solid lines show the posteriors obtained using a log-uniform prior on $\Omega_{\rm ref}$ and either a Gaussian (blue) or uniform (red) prior on $\alpha$. The priors and models are the same as those used in Ref.~\cite{O4a_iso_paper}.

\begin{table*}[t]
    \centering
    \begin{tabular}{c|c|c|c|c|c|c}
         & \multicolumn{3}{|c|}{\textbf{Uniform prior}} 
         & \multicolumn{3}{|c|}{\textbf{Log-uniform prior}}  \\ \hline
        \textbf{$\alpha$} & \textbf{O1--O4c1} & \textbf{O1--O4a} & \textbf{Improvement} 
                  & \textbf{O1--O4c1} & \textbf{O1--O4a} & \textbf{Improvement} \\ \hline \hline
        0 & \ULUnifAllZero & \ULUnifOoneOFourAZero & \ImprovementUnifZero 
          & \ULLogUnifAllZero & \ULLogUnifOoneOFourAZero & \ImprovementLogUnifZero   \\ 
        2/3  & \ULUnifAllTwoThirds & \ULUnifOoneOFourATwoThirds & \ImprovementUnifTwoThirds  
             & \ULLogUnifAllTwoThirds & \ULLogUnifOoneOFourATwoThirds & \ImprovementLogUnifTwoThirds  \\ 
        3 & \ULUnifAllThree & \ULUnifOoneOFourAThree & \ImprovementUnifThree 
          & \ULLogUnifAllThree & \ULLogUnifOoneOFourAThree & \ImprovementLogUnifThree  \\
        Marginalized & \ULUnifAllMarg & \ULUnifOoneOFourAMarg & \ImprovementUnifMarg 
                     & \ULLogUnifAllMarg & \ULLogUnifOoneOFourAMarg & \ImprovementLogUnifMarg \\ 
    \end{tabular}
    \caption{Bayesian upper limits at 95\% CL on the GWB amplitude $\Omega_{\rm ref}$ assuming uniform (left) and log-uniform (right) priors. Results obtained using O1--O4c1 data are compared with previous constraints from O1--O4a data. Limits are reported for fixed values of the spectral index $\alpha$ and for the case in which $\alpha$ is marginalized over assuming a Gaussian prior. The improvement relative to O1--O4a ranges from factors of 1.3 to 1.5 over the different priors and spectral indices. The O1--O4a upper limits differ from those in Ref.~\cite{O4a_iso_paper}, due to the updated notch list, as described in Sec.~\ref{sec:data_quality}.
} 
    \label{tab:upper_limits_Omega_ref}
\end{table*}

For a log-uniform prior on $\Omega_{\rm ref}$ (see Eq.~(\ref{power_law})), the log-Bayes factors comparing the signal and noise-only hypotheses are $\log_{10} \mathcal{B} = -0.518\pm 0.009$ and $\log_{10} \mathcal{B} =-0.573\pm 0.010$ for Gaussian and uniform priors on the spectral index $\alpha$, respectively. These values indicate a preference for the noise-only hypothesis. The one-dimensional posterior for $\Omega_{\rm ref}$ shows no significant dependence on the choice of prior for $\alpha$. Finally, the posterior for $\alpha$ no longer exhibits the peak around $\alpha = 5$ reported in Ref.~\cite{O4a_iso_paper}. This change arises from the investigations described in Sec.~\ref{sec:data_quality}, which identified additional noise features—now included in the notch list—that were responsible for the previously observed peak.

As noted above, the upper limits from O1--O4a reported in Table~\ref{tab:upper_limits_Omega_ref} differ from those in Ref.~\cite{O4a_iso_paper} due to the adoption of a new notch list. While the updated frequency-domain cuts described in Sec.~\ref{sec:data_quality} do not substantially affect the O1--O4a sensitivity for spectral indices $\alpha = 0,\, 2/3,\, 3$, they do modify the broadband estimator values from the inverse-variance-weighting over the new set of frequencies. This leads to more conservative upper limits compared to those in Ref.~\cite{O4a_iso_paper} for the spectral indices $\alpha = 0$ and $2/3$.

We find that the upper limits from O1--O4c1 are more stringent than those from O1--O4a by a factor of 1.3 to 1.5 for both uniform and log-uniform priors on $\Omega_{\rm ref}$.

\subsection{Non-GR polarizations}
\label{sec:non_GR}
Following the Bayesian framework adopted in Ref.~\cite{O4a_iso_paper}, and based on the formalism introduced in Ref.~\cite{Callister:2017ocg}, we perform a search aimed at improving constraints on possible deviations from GR~\cite{PhysRevD.79.082002}. The detection of non-tensorial polarization modes would provide direct evidence for a violation of GR, while their non-observation enables us to place limits on extended theories of gravity.

We adopt different models corresponding to GWBs with different polarization content. In particular, we first consider a mixed-polarization background, as described in Sec.~II~C of Ref.~\cite{O4a_iso_paper}, allowing for tensor, vector, and scalar components. We adopt log-uniform priors on the amplitude parameters $\Omega_{\mathrm{ref,(p)}}$, with $\mathrm{(p) \in \{T, V, S\}}$, over the range $10^{-13}–10^{-6}$, and Gaussian priors on the spectral indices $\alpha_{\mathrm{ref,(p)}}$, with $\mathrm{(p) \in \{T, V, S\}}$, centered at zero with a standard deviation of 3.5. From this analysis, we derive 95\% credible level upper limits on $\Omega_{\mathrm{ref,(p)}}$, summarized in the second column of Table~\ref{tab:UL_non_GR_pols}.

\begin{table*}[t]
%\vspace{1em} % adds breathing space
    \centering
    \begin{tabular}{c|c|c|c|c}
        \textbf{Polarization} & \textbf{O1$-$O4c1} & \textbf{O1$-$O4a (updated)} &  \textbf{Improvement} \\ \hline \hline
        Tensor &   \ULNonGRTVSTNew & \ULNonGRTVSTOld &   \ImprovementULNonGRTVST \\ \hline
        Vector &   \ULNonGRTVSVNew & \ULNonGRTVSVOld &   \ImprovementULNonGRTVSV \\ \hline
        Scalar &   \ULNonGRTVSSNew & \ULNonGRTVSSOld &   \ImprovementULNonGRTVSS \\ 
    \end{tabular}
    \caption{95\% CL upper limits on the amplitude of each polarization $\rm (p)$ component ($\Omega_{\mathrm{ref,p}}(25\,\mathrm{Hz})$) for a mixed-polarization background. Relative to the updated O1--O4a results, which supersede those reported in Ref.~\cite{O4a_iso_paper}, the limits improve by factors of 1.3--1.5. The upper limit on the scalar component remains less stringent because of its smaller overlap reduction function~\cite{Callister:2017ocg}.}
    \label{tab:UL_non_GR_pols}
%\vspace{1em} % adds breathing space
\end{table*}

As described previously, the upper limits reported in Ref.~\cite{O4a_iso_paper} are not directly comparable to the updated results presented here. We therefore recompute the O1--O4a upper limits using the stricter notch list described in Sec.~\ref{sec:data_quality}, obtaining the values reported in column~3 of Table~\ref{tab:UL_non_GR_pols}. This yields improvements in the O1--O4c1 upper limits, with respect to the recomputed O1–O4a results, by factors of $1.5$, $1.4$, and $1.3$ for the tensor, vector, and scalar polarizations, respectively.

The upper limit for the scalar-polarized case remains less stringent because the scalar overlap reduction function is approximately three times smaller in amplitude than the corresponding tensor and vector functions~\cite{Callister:2017ocg}. Consequently, the detector network is less sensitive to a scalar-polarized GWB.

We also perform dedicated searches under the assumption of purely tensor, vector, and scalar backgrounds, obtaining 95\% credible upper limits of $\Omega_{\mathrm{ref,T}}(25\,\mathrm{Hz}) \leq \ULNonGRT$, $\Omega_{\mathrm{ref,V}}(25\,\mathrm{Hz}) \leq \ULNonGRV$, and $\Omega_{\mathrm{ref,S}}(25\,\mathrm{Hz}) \leq \ULNonGRS$, respectively, when marginalizing over the spectral index. Relative to the updated O4a results, these correspond to improvements by factors of $1.5$, $1.3$, and $1.2$ for the tensor, vector, and scalar polarizations. 

As explained in Ref.~\cite{O4a_iso_paper}, the upper limits remain comparable across polarization because the O4 analysis uses only the LHO--LLO baseline, unlike O3 where the inclusion of Virgo data improved the ability to distinguish among different polarization modes.

To quantify the evidence for a stochastic signal, we compute the Bayes factor comparing the signal and noise hypotheses as $\frac{1}{7} \sum_\mathcal{H} \mathcal{B}^{\mathcal{H}}_{\rm N}$ for $\mathcal{H} \in {\rm \{T, V, S, TV, TS, VS, TVS\}}$ obtaining $\log_{10} \mathcal{B} = \BayesSignalvsNoise$ at 68\% credibility (see Appendix C of Ref.~\cite{Callister:2017ocg}), which indicates no statistically significant evidence for a detection. Here, the signal hypothesis is defined as an equal-weighted mixture of the tensor, vector, and scalar components, as well as their pairwise and triply combined configurations (TV, TS, VS, and TVS), thereby treating all polarization scenarios on an equal footing.
Finally, we evaluate the Bayes factor comparing non-GR and GR-consistent polarization models yielding $\log_{10} \mathcal{B} = \BayesNonGRvsGR$ at 68\% credibility. This result indicates a preference for a tensor-polarized background, as predicted by GR, over vector, scalar, and mixed-polarization alternatives. However, in the absence of a detection, this preference is expected to be dominated by the Occam penalty associated with the additional unconstrained parameters introduced in the more complex models.

\subsection{Joint magnetic and stochastic analysis}
\label{sec:joint_mag_stoch}
To ensure that the effect of the contamination from  correlated magnetic noise is negligible on the search for a GWB,
we provide a complementary test by performing a Bayesian analysis with the O4b and O4c1 data on a joint model composed of correlated magnetic noise and a GWB signal described by a power law~(PL).
Following the methodology of Ref.~\cite{PhysRevD.102.102005}, we compute the CSD of the magnetic channels between the two detectors. The inside-to-strain transfer functions (see Ref.~\cite{O4a_iso_paper} for a detailed description of this transfer function) are modelled using a PL approximation. These components are then combined to estimate the equivalent fractional energy density associated with correlated magnetic noise.
 
Since we only include LHO and LLO data, the amplitudes $\kappa_{H,\,L}$ and spectral indices $\beta_{H,\,L}$ of the coupling functions described in Eq.~(20) of Ref.~\cite{O4a_iso_paper} are degenerate. Therefore, to avoid the degeneracy in the Bayesian analysis, we adopt the amplitude product $\kappa_{HL}=\kappa_H\kappa_L$ and index sum $\beta_{HL}=\beta_H+\beta_L$ as the non-degenerate parameters of the correlated magnetic noise model.

In contrast to the magnetic budget, which relies on measured transfer functions from the magnetic field to the strain channel, this additional analysis aims to estimate the impact of correlated magnetic noise using agnostic priors for $\kappa_{HL}$ and $\beta_{HL}$ that are independent of transfer function measurements.
Instead of setting priors driven by such measured transfer functions as in the O3 and O4a analyses~\cite{PhysRevD.104.022004,O4a_iso_paper}, 
we have substantially expanded the range of the prior distributions for $\kappa_{HL}$ and $\beta_{HL}$ compared with those adopted in the previous analysis to cover the detection sensitivity shown by the O1--O4a PI curve, 
thereby making this estimation more conservative.
Consequently, we adopt a log-uniform prior distribution,
ranging from $10^{-56}$ to $10^{-38}\,{\mathrm{pT}}^{-2}$,
for $\kappa_{HL}$ and a uniform prior from $-4$ to $22$ for $\beta_{HL}$~\cite{PhysRevD.104.022004}.

We estimate the evidence for a magnetic noise only model and a joint model of magnetic noise and a PL GWB signal.
The Bayes factor comparing the magnetic noise only model to the uncorrelated Gaussian noise model is $\log_{10}\mathcal{B}_N^{\text{MAG}}=\LogTenbfSchuNde \pm \LogTenbfSchuNdeErr$,
which indicates the absence of correlated magnetic noise in the narrow band point estimate spectra.
The Bayes factor comparing the joint model of magnetic noise and a PL GWB signal to the uncorrelated Gaussian noise model is $\log_{10}\mathcal{B}_N^{\text{PL+MAG}}=\LogTenbfSchuNdePLGaussAlpha \pm \LogTenbfSchuNdePLGaussAlphaErr$,
verifying the null result for detecting both the PL GWB signal and correlated magnetic noise.

%%%%%%%%%%%%%%%%%%%%%%%%%%%
\section{Astrophysical implications}
\label{sec:astro_implications}
\begin{figure*}
    \centering
    \begin{minipage}{0.48\textwidth}
        \centering
        \includegraphics[width =1\linewidth]{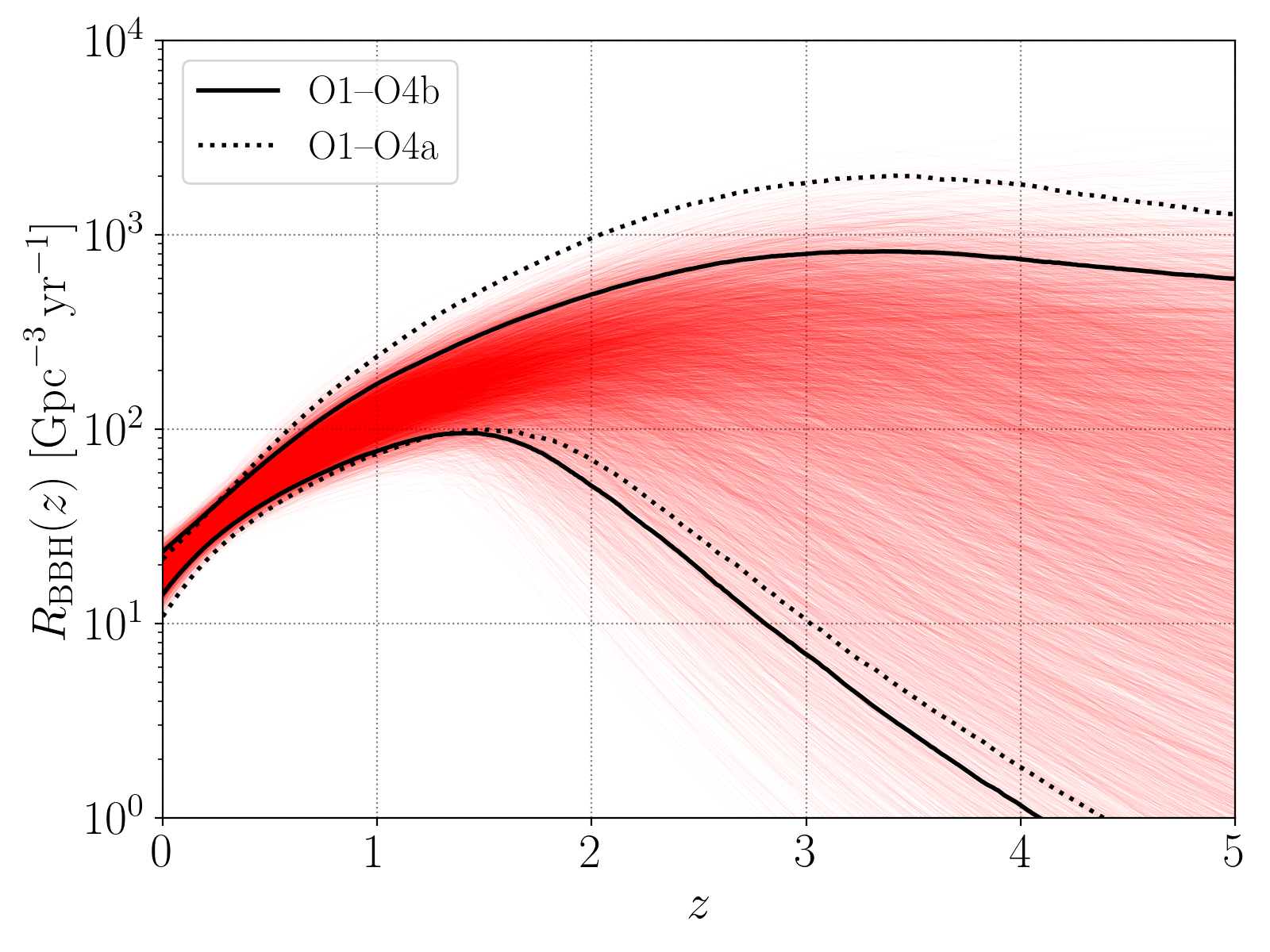}
    \end{minipage}\hfill
    \begin{minipage}{0.48\textwidth}
        \centering
        \includegraphics[width =1\linewidth]{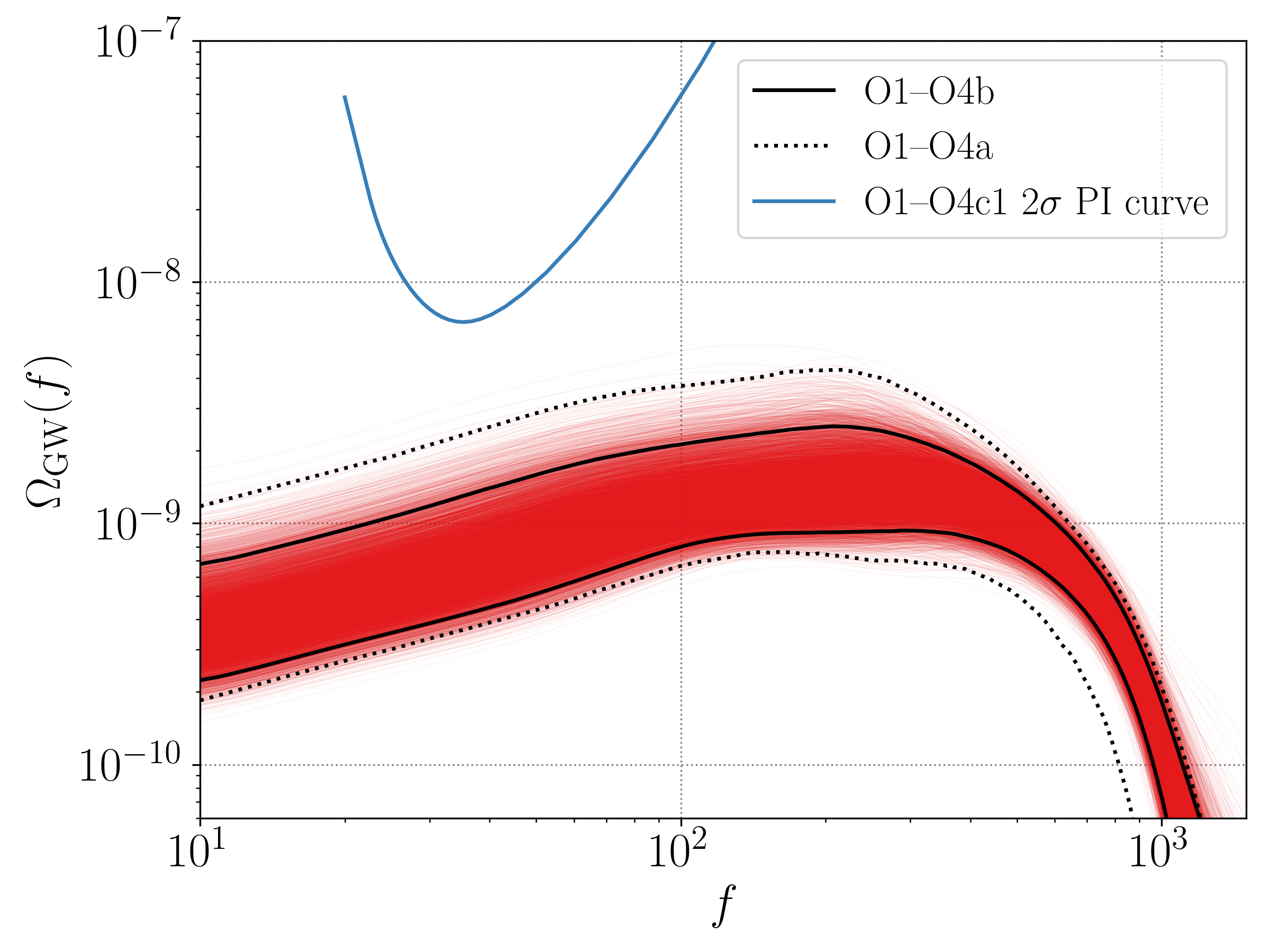}
    \end{minipage}
    \caption{\textit{Left panel:} Redshift evolution of the BBH merger rate density, inferred hierarchically from the BBH detections in GWTC-5.0, spanning observing runs O1--O4b. The red traces show the individual posterior samples, while the solid black lines indicate the 90\% credible bounds. For comparison, the 90\% credible bounds obtained previously with GWTC-4.0, which includes events through O4a, are shown as dotted lines.
    \textit{Right panel:} Prediction of the stochastic energy-density spectrum of the BBH population. The red traces correspond to the individual posterior samples on the energy-density spectrum obtained by sampling over the possible mass, spin and redshift distributions of BBH mergers, while the solid black curves indicate 90\% credible bounds on $\Omega_\mathrm{BBH}(f)$. For comparison, we also indicate the 90\% credible bounds obtained after O4a with dotted lines as presented in Ref.~\cite{O4a_iso_paper}. Furthermore, we show the $2\sigma$ PI sensitivity curve from O1--O4c1 data.}
    \label{fig:bbhforecast}
\end{figure*}

To interpret our upper limits, we place them in the context of the astrophysical GWB expected from the population of compact binary mergers inferred from the most recent LVK catalog, GWTC‑5.0~\cite{GWTC-5-Catalog:2026}.
Note that this analysis includes GW events detected up to the end of O4b, as a GWTC-6.0 catalog incorporating O4c events is not yet available.
In the frequency band of interest, the CBC background is expected to dominate the isotropic GWB. Following the O4a update~\cite{O4a_iso_paper}, we compute the contributions from BBH, BNS and NSBH systems at the 90\% credible level based on recent GW observations.
For the BBH population, we perform hierarchical Bayesian population inference and propagate the inferred hyperposterior into predictions for $\Omega_\mathrm{GW}(f)$. We adopt the same strongly parameterized population model used in Refs.~\cite{GWTC-4-Population,GWTC-5-Population:2026}. 
In this model, the primary mass distribution is described by a broken PL between $3$ and $200 \, M_\odot$ with two Gaussian peaks at $\sim 10\,M_\odot$ and $\sim 35\,M_\odot$, and tapering applied at the low-mass end to ensure a smooth turn-on of the distribution. The mass ratio follows a PL, spin magnitudes are drawn from a truncated normal distribution, and spin tilts are modeled as a mixture of isotropic and preferentially aligned configurations. As in Ref.~\cite{O4a_iso_paper}, but in contrast to the PL redshift evolution assumed in Ref.~\cite{GWTC-5-Population:2026}, we employ a generic broken PL redshift model that allows for a turnover in the merger rate history,
    \begin{equation}
    R_{\rm BBH}(z) = C(\alpha_z, \beta_z, z_{\rm p})
        \frac{ R_0 (1+z)^{\alpha_z}}{1 + \left(\frac{1+z}{1+z_{\rm p}}\right)^{\alpha_z+\beta_z}}\,,
    \label{eq:r-z}
    \end{equation}
where $\alpha_z$ and $\beta_z$ characterize the low- and high-redshift slopes, respectively, and $z_{\rm p}$ denotes the turnover redshift at which the merger rate density transitions from its low-redshift rise to its high-redshift decline. The normalization constant $C(\alpha_z,\beta_z,z_{\rm p})$ ensures that the merger rate density is normalized to the local rate $R_{\rm BBH}(0)=R_0$.
This redshift model is motivated by the fact that high-redshift sources can contribute substantially to the GWB and theoretical models of the merger rate evolution generally predict a turnover~\cite{Zhu:2011,Dominik:2013,Boesky:2024}.
Using the same priors on the mass and spin distributions as in Ref.~\cite{GWTC-5-Population:2026}, together with the redshift‑evolution priors as in the O4a analysis~\cite{O4a_iso_paper}, we recover mass and spin distributions that are fully consistent with the GWTC‑5.0 population study~\cite{GWTC-5-Population:2026}. We infer a local BBH merger rate of $R_0 = \RBBHMedianOFourb^{+\RBBHUpperErrorOFourb}_{-\RBBHLowerErrorOFourb}\, \mathrm{Gpc}^{-3}\mathrm{yr}^{-1}$, which is likewise consistent with the results reported in Ref.~\cite{GWTC-5-Population:2026}, noting that the BBH merger rates quoted therein are evaluated at $z=0.2$.
While the high‑redshift slope $\beta_z$ remains unconstrained, we obtain positive values for the low‑redshift index $\alpha_z = \alphaMedianOFourb^{+\alphaUpperErrorOFourb}_{-\alphaLowerErrorOFourb}$, which are somewhat lower than those reported in the O4a analysis. We also find a mild preference for turnover redshifts $z_\mathrm{p} \simeq 2$, with very low values of $z_\mathrm{p}$ disfavored. 
This shift towards a more moderate increase of the BBH merger rate with redshift, also consistently observed in Ref.~\cite{GWTC-5-Population:2026}, results in a credible band that is overall shifted to lower merger rates compared to O4a, as illustrated in Fig.~\ref{fig:bbhforecast}.

Following the procedure outlined in Refs.~\cite{O4a_iso_paper,Renzini:2024,Ebersold:2025}, we compute the fractional energy density spectrum for each posterior sample of the BBH population model. In the absence of a stochastic detection, we apply the rejection sampling method of Ref.~\cite{O4a_iso_paper} to combine constraints from direct BBH detections with those from the stochastic search~\cite{Callister:2020}. 
The individual $\Omega_\mathrm{BBH}(f)$ spectra are retained or rejected based on their consistency with the measured cross-correlation data.
Because the majority of the spectra lie well below the search upper limits, only a small fraction of samples is rejected. The resulting ensemble of accepted spectra, together with the corresponding 90\% credible band, is shown in Fig.~\ref{fig:bbhforecast}. 
At the reference frequency of $25\,\mathrm{Hz}$, we obtain a predicted BBH background amplitude of $\Omega_\mathrm{BBH}(25 \,\mathrm{Hz}) = \bbhMedian^{+\bbhUpperError}_{-\bbhLowerError} \times 10^{\bbhPower}$. This represents a substantial improvement over previous constraints: although the lower bound remains similar, the upper bound is reduced by almost a factor of two. This tightening primarily arises from the improved constraints on the BBH merger-rate density, in particular on the low-redshift slope $\alpha_z$.
Looking ahead, we expect the growing catalog of directly detected BBH mergers to further tighten constraints on $R_0$, $\alpha_z$ and the low-redshift side of $z_\mathrm{p}$. In contrast, stochastic background measurements will provide complementary information on the merger rate density at higher redshifts, constraining $\beta_z$ and the high-redshift side of $z_\mathrm{p}$.

Beyond the BBH contribution, we also predict the GWB generated by BNS and NSBH systems, following the same procedure as in Ref.~\cite{O4a_iso_paper}. Owing to the limited constraints from direct observations of these populations, we estimate their local rates under fixed assumptions about their mass and spin distributions, and we model their redshift evolution by convolving the cosmic star-formation history with a distribution of delay times between formation and merger. Specifically, we assume that progenitor formation traces the star-formation rate density of Ref.~\cite{Madau:2016} and that delay times are distributed log-uniformly between $20\,\mathrm{Myr}$ and $13.5\,\mathrm{Gyr}$.
For the BNS population, we adopt a uniform component mass distribution between $1$ and $2.5\,M_\odot$ and isotropic spin orientations with spin magnitudes uniformly distributed below 0.4.
Under these assumptions, we infer a local BNS merger rate of $ \bnsrateMedian^{+\bnsrateUpperError}_{-\bnsrateLowerError} \,\mathrm{Gpc}^{-3}\mathrm{yr}^{-1}$. This estimate is consistent with the BNS merger rate reported in Ref.~\cite{GWTC-5-Population:2026}, although that analysis jointly inferred the BNS and BBH merger rates across the full compact binary mass range.
Using this merger rate, we predict a BNS background amplitude of $\Omega_\mathrm{BNS}(25 \,\mathrm{Hz}) = \bnsMedian^{+\bnsUpperError}_{-\bnsLowerError} \times 10^{\bnsPower}$, with the dominant uncertainty arising from the local merger rate.

\begin{figure}[h]
    \centering
    \includegraphics[width=1\linewidth]{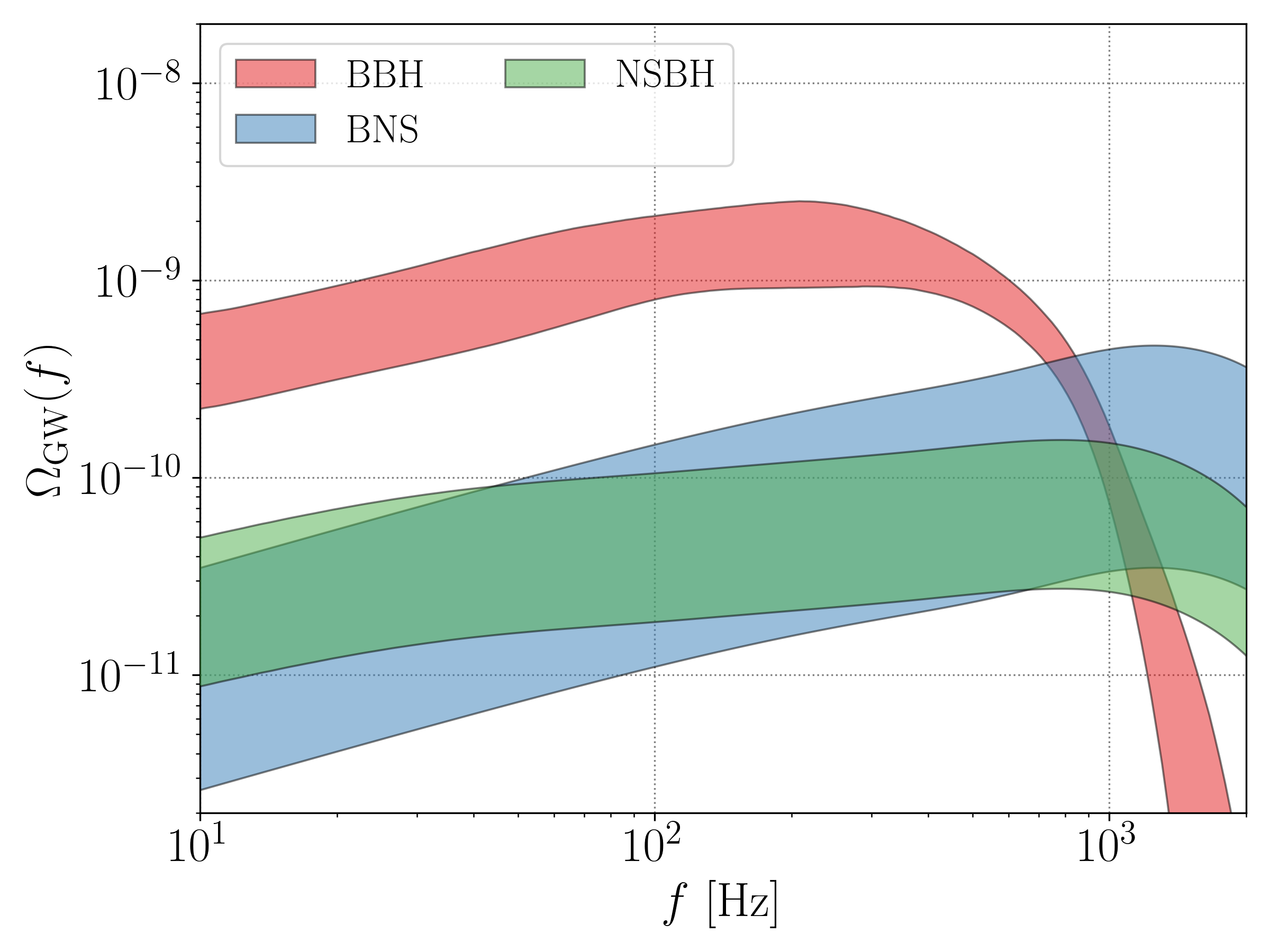}
    \caption{Predicted astrophysical GWBs from BBH, BNS and NSBH coalescences at the 90\% credible level.}
    \label{fig:backgrounds}
\end{figure}

For NSBH systems, we assume that the neutron-star component follows the same mass and spin distributions as in the BNS population, while the black-hole component is drawn from a log-uniform mass distribution between $3$ and $50\,M_\odot$, with isotropically oriented spins and magnitudes uniformly distributed below 0.99. These assumptions yield a local NSBH merger rate of $\nsbhrateMedian^{+\nsbhrateUpperError}_{-\nsbhrateLowerError} \,\mathrm{Gpc}^{-3}\mathrm{yr}^{-1}$, and a corresponding stochastic background amplitude of $\Omega_\mathrm{NSBH}(25 \,\mathrm{Hz}) = \nsbhMedian^{+\nsbhUpperError}_{-\nsbhLowerError} \times 10^{\nsbhPower}$.
Figure~\ref{fig:backgrounds} shows the resulting 90\% credible intervals for the BNS and NSBH backgrounds alongside the BBH contribution. For both BNS and NSBH, the credible bands narrow relative to earlier estimates, primarily due to the improved constraints on their local merger rates.

\begin{figure}[h]
    \centering
    \includegraphics[width =1\linewidth]{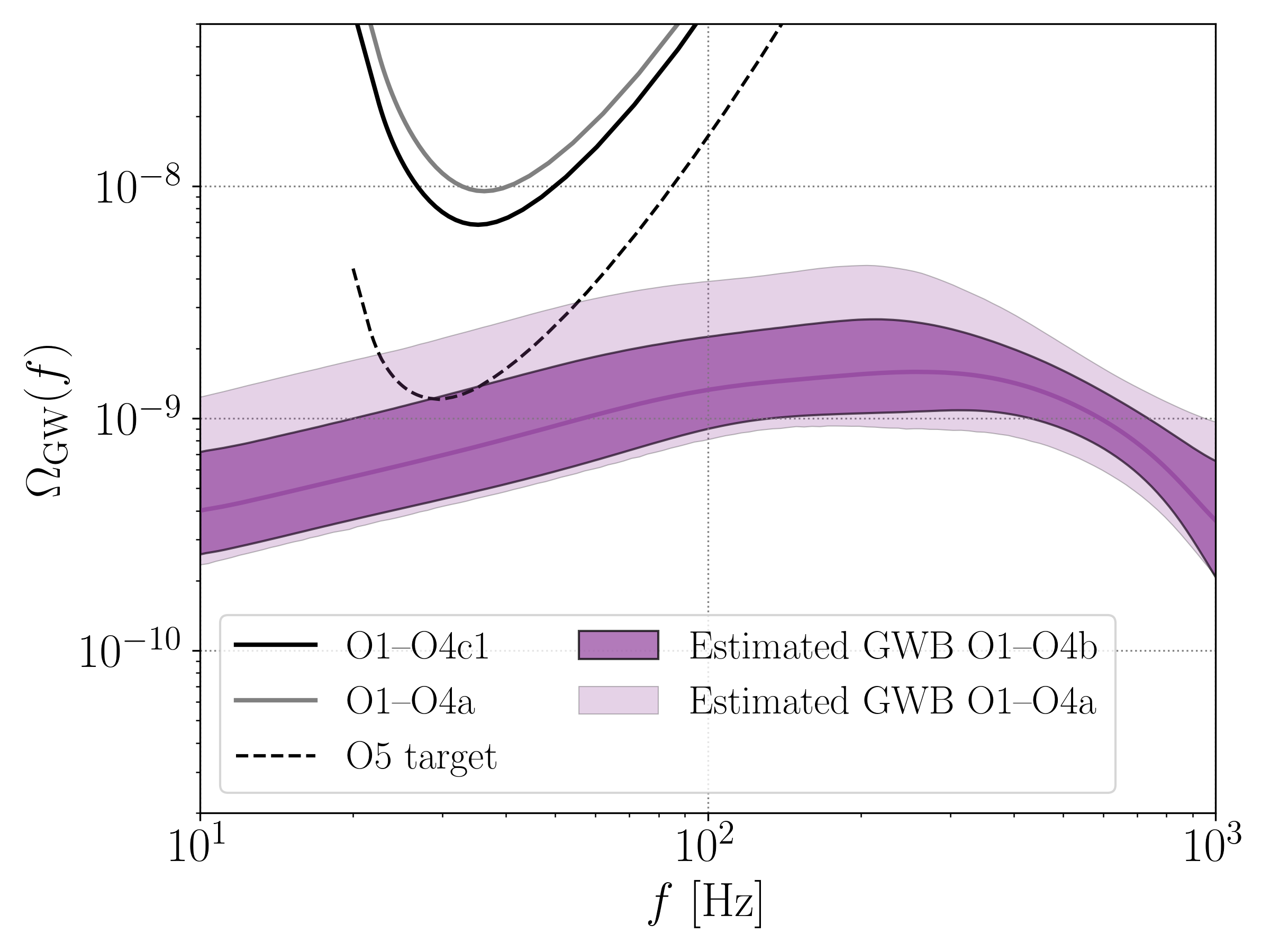}
    \caption{The purple band shows the combined estimate of the CBC background at the 90\% credible level, with the median indicated by the solid line. For comparison, the previous estimate after O4a from Ref.~\cite{O4a_iso_paper} is also shown. The figure further includes the $2\sigma$ PI sensitivity curve of the O1--O4c1 search presented in this work, together with the corresponding curve from the O4a search and the projected target sensitivity for one year of coincident O5 data. The O5 projection assumes equal observing time at the expected O5a, O5b and O5c sensitivity levels.}
    \label{fig:cbcbackground}
\end{figure}

We also combine the individual BBH, BNS and NSBH forecasts to obtain a total CBC background estimate, for which we find $\Omega_\mathrm{CBC}(25 \,\mathrm{Hz}) = \cbcMedian^{+\cbcUpperError}_{-\cbcLowerError} \times 10^{\cbcPower}$. The upper bound of this credible interval is roughly a factor of two lower than the value reported in the previous LVK stochastic background search~\cite{O4a_iso_paper}, mainly due to the weaker rise of the BBH merger rate evolution at low redshift, as discussed above.
Figure~\ref{fig:cbcbackground} shows the combined estimate alongside the $2\sigma$ PI curve representing the integrated sensitivity of the search reported here, as well as the updated projected O5 sensitivity of the LIGO detectors~\cite{O5strain-projection}. Since the O5 sensitivity projection is provided for three representative epochs with progressively improving sensitivity levels, the O5 PI curve shown here is constructed assuming one year of coincident data, with one third of the observing time spent at each of the O5a, O5b and O5c sensitivity levels.
With the increased number of GW observations and the improved understanding of the local CBC populations, the uncertainty on our CBC background estimate is substantially reduced. Nevertheless, under the projected O5 sensitivity, a detection of the CBC background remains within reach, provided the detectors achieve their anticipated performance.

%%%%%%%%%%%%%%%%%%%%%%%%%%%%%%%%%%%%%%%%%%%%%%%
\section{Conclusions}
\label{sec:conclusions}
In this work, we performed a cross-correlation search for an isotropic gravitational-wave background using data from the first LVK observing run (O1) through April 1, 2025 (O4c1). Relative to the previous analysis in Ref.~\cite{O4a_iso_paper}, this search incorporated an additional 117 days of coincident LIGO Hanford--Livingston data from O4b and O4c1, after data-quality cuts. We did not include Virgo data from O4, as their addition improved the network sensitivity by less than 1\%. The analysis covered the frequency band from 20 to 1726~Hz.

During O4, new data-quality challenges emerged in the form of noise features not identified by existing vetoing methods. To address this, we developed a stricter notch-list procedure based on the temporal variability of the power spectral density, which allowed us to identify both narrow spectral lines and broader non-stationary features that remained hidden in time-averaged spectra. This produced two new notch lists: an updated one for O4a and another for O4b--O4c1, which made it necessary to reanalyze the O1--O4a data with the revised frequency cuts.

We found no evidence for an isotropic background. Hence, we placed a 95\% credible upper limit of e.g. $\Omega_{\mathrm{ref}}(25\,\mathrm{Hz}) \leq \ULLogUnifAllZero$ for a spectral index $\alpha=0$ under a log-uniform prior ($\leq \ULUnifAllZero$ under a uniform prior), improving the recalculated O1--O4a constraints by a factor of $\ImprovementLogUnifZero$ ($\ImprovementUnifZero$ for a uniform prior). 

Since cross-correlation searches could be biased by correlated environmental noise, we also evaluated the contribution from magnetic Schumann resonances and found that correlated magnetic noise did not affect the results in this paper. Furthermore, we performed a joint Bayesian search in which correlated magnetic noise and the gravitational-wave background were assumed to be present simultaneously, which showed no evidence for such a scenario.

We also searched for alternative gravitational-wave polarizations beyond the tensor modes predicted by general relativity, namely scalar and vector components. We improved O1--O4a constraints on these non-standard polarization backgrounds by factors of $\ImprovementULNonGRTVSV$ and $\ImprovementULNonGRTVSS$ for vector and scalar modes, respectively. Our analysis indicated a preference for a tensor-polarized background, as predicted by GR. However, in the absence of a detection, this apparent preference is expected to be driven primarily by the Occam penalty associated with the additional unconstrained parameters of the more complex polarization models.

We updated the forecast of the GWB from compact binaries using population inferences based on GWTC-5.0. By propagating uncertainties in the population models, we obtained refined estimates for the BBH, BNS and NSBH contributions and combined them into an updated prediction for the total CBC background. 
The inferred background amplitude is lower and more tightly constrained than in previous analyses, primarily due to improved measurements of the BBH merger-rate evolution, including evidence for a more moderate increase with redshift. While current observational limits remain above the predicted astrophysical background, the expected sensitivity improvements in O5 make a detection of the CBC background increasingly plausible, positioning stochastic searches as a complementary probe of compact binary formation and merger history across cosmic time.

%%%%%%%%%%%%%%%%%%%%%%%%%%%%%%%%%%%%%%%%%%%%%%%
\section*{ACKNOWLEDGMENTS}
\label{sec:acknowledgements}
This material is based upon work supported by NSF's LIGO Laboratory, which is a
major facility fully funded by the National Science Foundation.
This publication has been assigned LIGO document number P2600217.
The authors also gratefully acknowledge the support of
the Science and Technology Facilities Council (STFC) of the
United Kingdom, the Max-Planck-Society (MPS), and the State of
Niedersachsen/Germany for support of the construction of Advanced LIGO 
and construction and operation of the GEO\,600 detector. 
Additional support for Advanced LIGO was provided by the Australian Research Council.
The authors gratefully acknowledge the Italian Istituto Nazionale di Fisica Nucleare (INFN),  
the French Centre National de la Recherche Scientifique (CNRS) and
the Netherlands Organization for Scientific Research (NWO)
for the construction and operation of the Virgo detector
and the creation and support  of the EGO consortium. 
The authors also gratefully acknowledge research support from these agencies as well as by 
the Council of Scientific and Industrial Research of India, 
the Department of Science and Technology, India,
the Science \& Engineering Research Board (SERB), India,
the Ministry of Human Resource Development, India,
the Spanish Agencia Estatal de Investigaci\'on (AEI),
the Spanish Ministerio de Ciencia, Innovaci\'on y Universidades,
the European Union NextGenerationEU/PRTR (PRTR-C17.I1),
the ICSC - CentroNazionale di Ricerca in High Performance Computing, Big Data
and Quantum Computing, funded by the European Union NextGenerationEU,
the Comunitat Auton\`oma de les Illes Balears through the Conselleria d'Educaci\'o i Universitats,
the Conselleria d'Innovaci\'o, Universitats, Ci\`encia i Societat Digital de la Generalitat Valenciana and
the CERCA Programme Generalitat de Catalunya, Spain,
the Polish National Agency for Academic Exchange,
the National Science Centre of Poland and the European Union - European Regional
Development Fund;
the Foundation for Polish Science (FNP),
the Polish Ministry of Science and Higher Education,
the Swiss National Science Foundation (SNSF),
the Russian Science Foundation,
the European Commission,
the European Social Funds (ESF),
the European Regional Development Funds (ERDF),
the Royal Society, 
the Scottish Funding Council, 
the Scottish Universities Physics Alliance, 
the Hungarian Scientific Research Fund (OTKA),
the French Lyon Institute of Origins (LIO),
the Belgian Fonds de la Recherche Scientifique (FRS-FNRS), 
Actions de Recherche Concert\'ees (ARC) and
Fonds Wetenschappelijk Onderzoek - Vlaanderen (FWO), Belgium,
the Paris \^{I}le-de-France Region, 
the National Research, Development and Innovation Office of Hungary (NKFIH), 
the National Research Foundation of Korea,
the Natural Sciences and Engineering Research Council of Canada (NSERC),
the Canadian Foundation for Innovation (CFI),
the Brazilian Ministry of Science, Technology, and Innovations,
the International Center for Theoretical Physics South American Institute for Fundamental Research (ICTP-SAIFR), 
the Research Grants Council of Hong Kong,
the National Natural Science Foundation of China (NSFC),
the Israel Science Foundation (ISF),
the US-Israel Binational Science Fund (BSF),
the Leverhulme Trust, 
the Research Corporation,
the National Science and Technology Council (NSTC), Taiwan,
the United States Department of Energy,
and
the Kavli Foundation.
The authors gratefully acknowledge the support of the NSF, STFC, INFN and CNRS for provision of computational resources.
This work was supported by MEXT,
the JSPS Leading-edge Research Infrastructure Program,
JSPS Grant-in-Aid for Specially Promoted Research 26000005,
JSPS Grant-in-Aid for Scientific Research on Innovative Areas 2402: 24103006,
24103005, and 2905: JP17H06358, JP17H06361 and JP17H06364,
JSPS Core-to-Core Program A.\ Advanced Research Networks,
JSPS Grants-in-Aid for Scientific Research (S) 17H06133 and 20H05639,
JSPS Grant-in-Aid for Transformative Research Areas (A) 20A203: JP20H05854,
the joint research program of the Institute for Cosmic Ray Research,
University of Tokyo,
the National Research Foundation (NRF),
the Computing Infrastructure Project of the Global Science experimental Data hub
Center (GSDC) at KISTI,
the Korea Astronomy and Space Science Institute (KASI),
the Ministry of Science and ICT (MSIT) in Korea,
Academia Sinica (AS),
the AS Grid Center (ASGC) and the National Science and Technology Council (NSTC)
in Taiwan under grants including the Science Vanguard Research Program,
the Advanced Technology Center (ATC) of NAOJ, 
the Mechanical Engineering Center of KEK
and Vietnam National Foundation for Science and Technology Development 
(NAFOSTED) 103.01-2025.147.
Additional acknowledgements for support of individual authors may be found in the following document: \\
\texttt{https://dcc.ligo.org/LIGO-M2300033/public}.
For the purpose of open access, the authors have applied a Creative Commons Attribution (CC BY)
license to any Author Accepted Manuscript version arising.
% Update next sentence to another first author in the future, as needed
We request that citations to this article use 'A. G. Abac {\it et al.} (LIGO-Virgo-KAGRA Collaboration), ...' or similar phrasing, depending on journal convention.

%%%%%%%%%%%%%%%%%%%%%%%%%%%%%%%%%%%%%%%%%%%%%%%

\bibliographystyle{apsrev4-1}
\bibliography{References}
\end{document}